# Resonant tunneling in graphene pseudomagnetic quantum dots

Zenan Qi,[†,§] D. A. Bahamon,[‡,§] Vitor M. Pereira,[*,‡] Harold S. Park,[†] D. K. Campbell,[¶] and A. H. Castro Neto[‡,¶]

*Department of Mechanical Engineering, Boston University, Boston, MA 02215, Graphene Research Centre & Department of Physics, National University of Singapore, 2 Science Drive 3, Singapore 117542, and Department of Physics, Boston University, Boston, MA 02215, USA*

E-mail: vpereira@nus.edu.sg

---

*To whom correspondence should be addressed
†Department of Mechanical Engineering, Boston University, Boston, MA 02215
‡Graphene Research Centre & Department of Physics, National University of Singapore, 2 Science Drive 3, Singapore 117542
¶Department of Physics, Boston University, Boston, MA 02215, USA
§These authors contributed equally to this work.

**Abstract**

Realistic relaxed configurations of triaxially strained graphene quantum dots are obtained from unbiased atomistic mechanical simulations. The local electronic structure and quantum transport characteristics of y-junctions based on such dots are studied, revealing that the quasi-uniform pseudomagnetic field induced by strain restricts transport to Landau level- and edge state-assisted resonant tunneling. Valley degeneracy is broken in the presence of an external field, allowing the selective filtering of the valley and chirality of the states assisting in the resonant tunneling. Asymmetric strain conditions can be explored to select the exit channel of the y-junction.

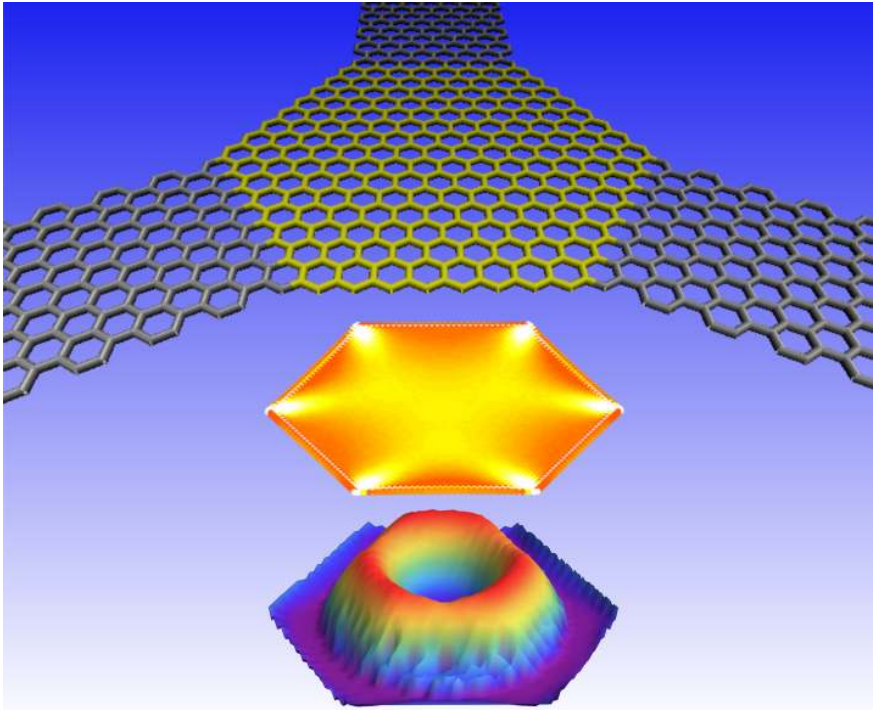

*Illustration of a strained y-junction with the associated pseudomagnetic field distribution and LDOS map of one of the pseudomagnetic edge modes contributing to the edge state-assisted resonant tunneling behavior of these structures.*

## Introduction

Endowed with the strongest covalent bonding in nature, graphene exhibits the largest tensional strength ever registered ($E \simeq 1\,\mathrm{TPa}$), and a record range of elastic deformation for a crystal, which can be as high as 15-20%.[1,2] Such outstanding mechanical characteristics are complemented by an unusual coupling of lattice deformations to the electronic motion, that can be captured by the concept of a pseudomagnetic field (PMF) arising as a result of non-uniform local changes in the electronic hopping amplitudes.[3–5] Since electrons in graphene respond to these local PMFs precisely as they would to a real magnetic field, this specific strain-induced perturbation is not screened by the free electrons in the same way that the usual displacement field coupling can be. Consequently, the ability to manipulate the strain distribution in graphene opens the enticing prospect of strain-engineering its electronic and optical properties, as well as of enhancing interaction and correlation effects.[6–11] The recent experimental confirmation that PMFs in excess of $300\,\mathrm{T}$ are possible with modest deformations in structures spanning only a few nm[12,13] brings this prospect of strongly impacting graphene's electronic properties by strain closer to fruition.

Despite this recent experimental evidence for strong PMF-induced Landau quantization, to the best of our knowledge, no measurements or calculations have been performed to assess the transport characteristics of such nano-structures. The ability to produce very large and fairly homogeneous PMFs within a few nm suggests the possibility of creating pseudomagnetic quantum dots, where confinement is driven by the PMF. Here we undertake a theoretical study to probe the electronic and quantum transport properties of a representative graphene-based strained y-junction particularly suited to the generation of quasi-uniform PMFs.[6] We assess the quantum transport characteristics of such structures, revealing the Landau level- (LL) assisted character of the tunneling mechanism, as well as the interplay with an external field that breaks the valley degeneracy.

## Calculation methodology

In order to capture the microscopic details as realistically as possible, we employed a combined atomistic, electronic and transport calculation procedure, which provides a set of unbiased results at all these levels. The microscopic configuration of each carbon atom is obtained by a fully relaxed Molecular Mechanics (MM) approach which, together with Monte Carlo approaches,[14] constitutes one of the most unbiased ways to describe deformation fields in nanostructures. Knowledge of the position of each atom allows us to extract the $\pi$-band bandstructure of the relaxed lattice via a tight-binding (TB) approach, as well as to calculate the quantum transport properties across the structure via a non-equilibrium Green's function (NEGF) approach. In this way one unveils the local electronic structure, from which we can extract, e.g., the local pseudo-magnetic fields (PMF) and local current distribution, without approximations, using a system with more than 6000 atoms.

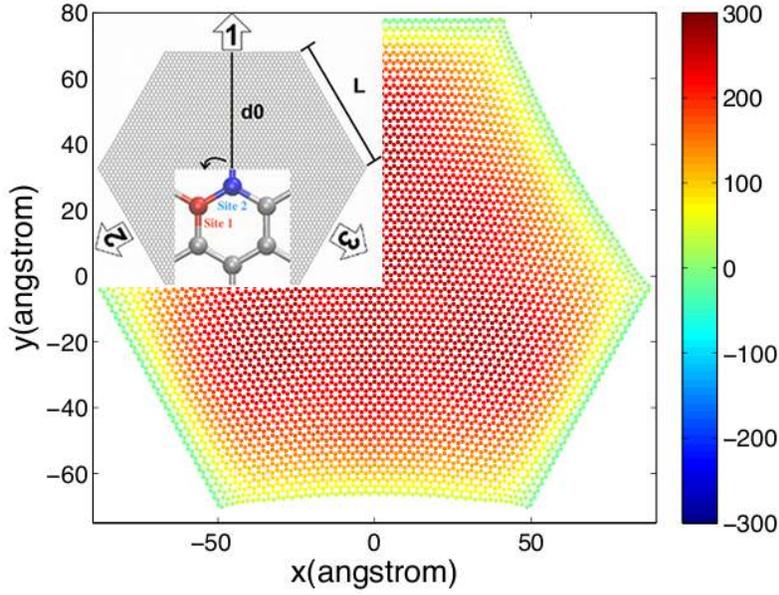

Figure 1: (Color online) Real-space distribution of the PMF $B_s$ (Tesla) under $\varepsilon_{\text{eff}} = 15\%$ obtained by mapping the tight-binding-derived LDOS at each atom. Inset: diagram of the tri-axial loading and contact scheme. $L \simeq 8\,\text{nm}$.

The deformed configurations of an hexagonal graphene monolayer were obtained using standard MM simulations at 0 K with LAMMPS.[15] For definiteness, we shall focus here on the system shown in Fig. 1, with 6144 atoms, and a side length, $L = 7.87$ nm, but we stress that the results do not show variance among specific sizes and can be straightforwardly rescaled to larger or smaller dimensions.[16,17] The carbon interactions were modeled via the AIREBO potential[18] with a cutoff at 0.68 nm, which has been shown to capture accurately the mechanical properties of carbon-based nanostructures, including bond breaking, deformation, and various elastic moduli.[19–21] Additional details are discussed in the Supporting Information (SI).[17] The system was triaxially stretched by in-plane displacement increments of $10^{-3}$ nm along each of the three arms shown in Fig. 1. Following each strain increment, graphene was allowed to relax according to the conjugate gradient algorithm, until relative changes in the system energy from one increment to the next were smaller than $10^{-7}$. Since the strain thus generated is non-uniform, we introduce the nominal strain $\varepsilon_{\text{eff}} = (d - d_0)/d_0$, where $d_0(d)$ is the distance from the center to the edge of the hexagon before (after) stretching, as illustrated in Fig. 1. Nominal strains ranging from 0 to 18% are considered below.

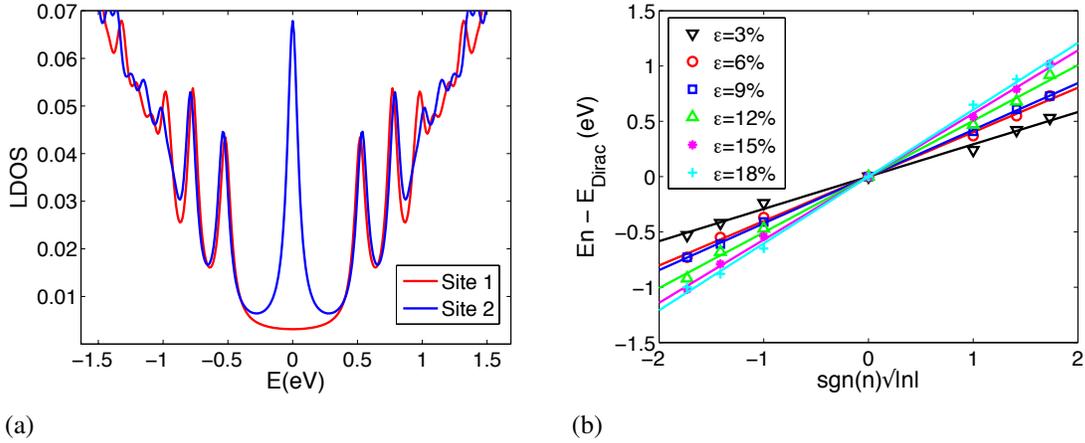

(a)

(b)

Figure 2: (Color online) (a) LDOS at two representative neighboring sites (Fig. 1) for $\varepsilon_{\text{eff}} = 15\%$. (b) Peak positions vs $\text{sgn}(n)\sqrt{|n|}$, extracted from spectra such as (a), and for different $\varepsilon_{\text{eff}}$. Straight lines are fits to Eq. (1) from which we extract the local $B_s$ at the site where the LDOS was sampled.

Once the relaxed configurations were obtained at each value of strain the atomic positions were used as the basis for electronic structure and quantum transport calculations. Specifically, we used the relaxed atomic positions as input to the exact diagonalization of the $\pi$-band TB Hamiltonian for graphene, using the parametrization $V_{pp\pi}(l) = t_0 e^{-3.37(l/a-1)}$ to describe the dependence of this Slater-Koster parameter on the C–C distance $l$ ($t_0 = 2.7$ eV and $a \simeq 0.142$ nm). This approximation was shown to describe with good accuracy both the threshold deformation for the Lifshitz band insulator transition at large deformations,[22,23] and the behavior of $V_{pp\pi}(l)$ or the optical conductiv-ity when directly compared to *ab-initio* calculations.[8,24] Here we do not consider electron-electron interactions.

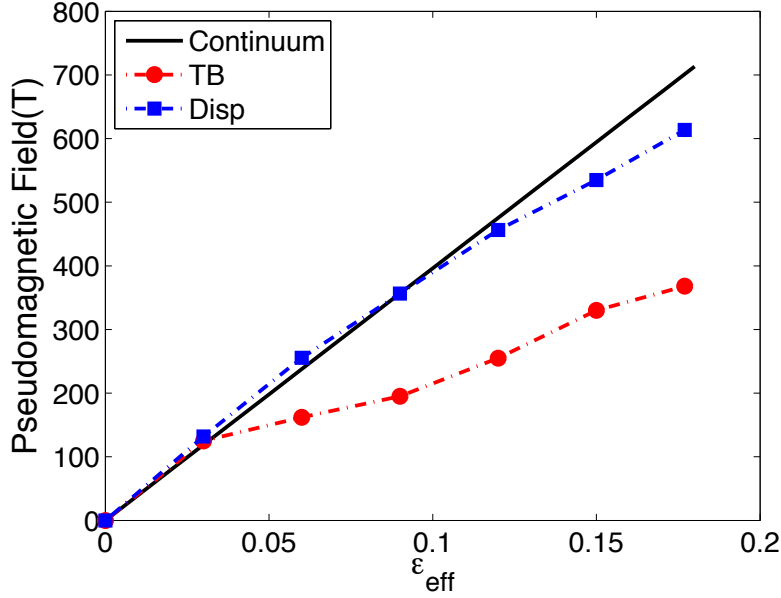

Figure 3: (Color online) Dependence of $B_s^{\max}$ on $\varepsilon_{\text{eff}}$ obtained by the tight binding and displacement approaches discussed in the text.

One property we extract from this procedure is the exact (within this TB model) local density of states (LDOS), from which we can map the local PMF distribution by fitting the resonant LDOS at each atom to the Landau level (LL) spectrum expected for graphene:[5,25]

$$E_n = \pm\hbar\omega_c\sqrt{n}, \quad \hbar^2\omega_c^2 = 2e\hbar v_F^2 B_s, \quad \hbar v_F = 3ta/2. \tag{1}$$

An example of typical LDOS spectra is shown in Fig. 2(a). Eq. (1) is used to obtain the local $B_s$ distribution throughout the system by fitting the slope of $E_n$ vs $\sqrt{n}$ seen in the numerical LDOS at various strains, as shown in Fig. 2(b). Notice how the $n = 0$ LL is absent in the LDOS of one of the sublattices, similarly to data recently reported in experiments with artificial honeycomb lattices.[26]

To complement this exact numerical calculation of $B_s$ at each lattice point, we used another approach for comparison and control. It hinges upon the strain-induced perturbation to the continuum Dirac equation that is applicable at low energies.[4,6,27] In this approach one uses $B_s = \partial_x A_y - \partial_y A_x$, where the fictitious vector potential for an electron (charge $-e$) is given by

$$A_x = -\frac{3t\beta}{4ev_F}(\varepsilon_{xx} - \varepsilon_{yy}), \qquad A_y = -\frac{3t\beta}{2ev_F}(-\varepsilon_{xy}), \tag{2}$$

and $\beta = -a\frac{\partial \ln V_{pp\pi}(l)}{\partial l} \approx 3.37$, the same used in our TB parametrization. The key here is to obtain the space-dependent strain tensor $\varepsilon_{ij}(\mathbf{r})$. It can be obtained directly from the displacements or the atomistic stresses during the MM calculation, and affords an alternative method to extract $B_s(\mathbf{r})$ in the entire system more expeditiously. We shall refer to this as the "displacement approach" and use both methods to map the PMF distribution, thus assessing the range of validity of this "displacement approach + Dirac equation" in comparison with the direct TB on the deformed lattice.[17]

## PMF distribution

Recent experiments show in a spectacular way how strain can impact the electronic properties of graphene by confirming the existence of strain-induced LLs corresponding to fields from 300 to 600 T in graphene nanobubbles.[12,13] Our approach of sampling the LDOS and fitting the LL resonances to Eq. (1), as illustrated in Fig. 2, is the theoretical analog of the STM analysis done in those experiments. The real space PMF distribution for $\varepsilon_{\text{eff}} = 15\%$ is shown in Fig. 1, and follows the general predictions of Guinea et al.[6]. Most notably, the PMF is nearly uniform in most of the inner portion, which is a consequence of the trigonal loading conditions. This field uniformity is crucial to have well defined LLs at nominal strains as small as 3%. To quantify the dependence of the field on the nominal strain we plot the maximum, $B_s^{\max}$ at the center of the hexagon in Fig. 3, showing that, for the parameter $\beta$ used here and at small $\varepsilon_{\text{eff}}$, each 1% of nominal strain increases $B_s^{\max}$ by $\approx 40$ T. Direct comparison of the curves generated by the two methods mentioned above shows that $B_s$ obtained using the "displacement approach" begins overestimating $B_s$ beyond $\varepsilon_{\text{eff}} \sim 5\%$. This is expected insofar as Eq. (2) results from an expansion of $V_{pp\pi}(l)$ to linear order in $\varepsilon_{\text{eff}}$, and hence is bound to overestimate the rate of change of $V_{pp\pi}$ (and thus $B_s$) at higher deformations. On the basis of our data at low $\varepsilon_{\text{eff}}$ we extract the scaling $B_s^{\max} = C \varepsilon_{\text{eff}}/L$, with $C \simeq 3 \times 10^4$ T nm. This relation can be used to obtain $B_s$ for systems with any $L$ and $\varepsilon_{\text{eff}}$. For large $\varepsilon_{\text{eff}}$ the data in Fig. 3 must be taken into account to correct for the overestimate. The magnetic lengths, $\ell_{B_s}$, associated with these large PMFs can easily become comparable to the system size, and thus a strong interplay between magnetic and spatial confinement is expected. In particular, the small size of the quantum dot implies that most low energy states will not be "condensed" into LLs[28] and, in addition, resonant transport behavior can be seen in these structures as a result of tunneling assisted by the magnetically confined states in the central region. This is characterized next.

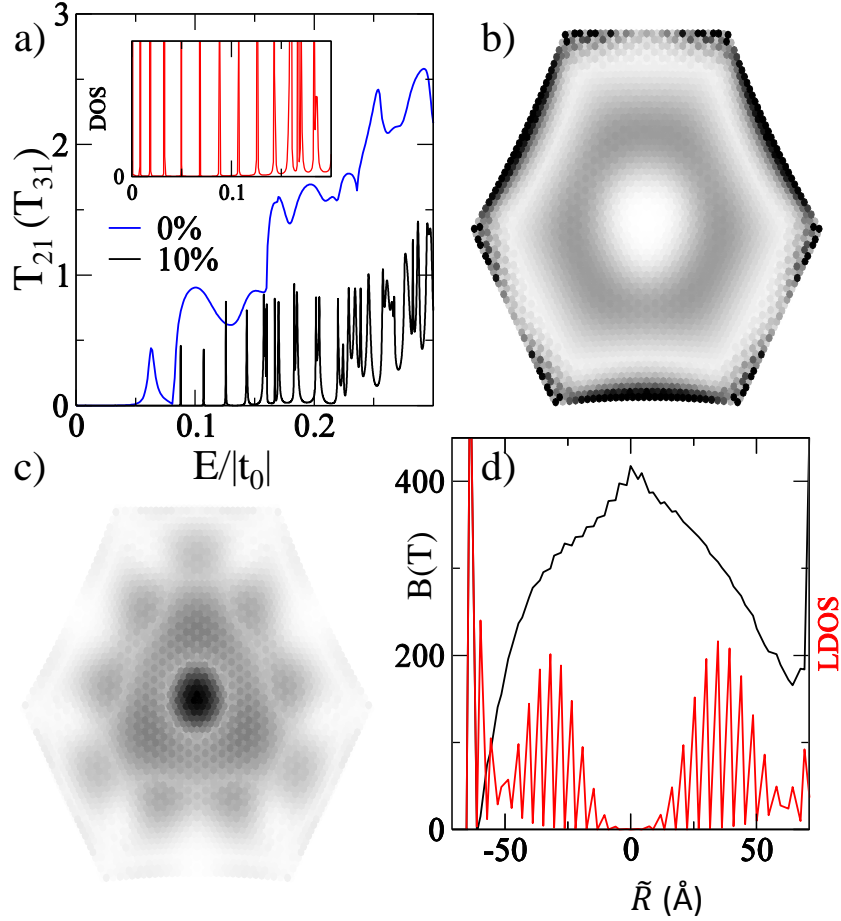

Figure 4: (Color online) (a) Transmission coefficient $T_{21}(T_{31})$ vs $E_F$ for $\varepsilon_{\text{eff}} = 0$ and 10%. The inset shows a close-up of the total DOS of the strained dot in the low energy region. A LDOS map (white is zero) of selected transmission resonances for $\varepsilon_{\text{eff}} = 10\%$ is shown in (b) for $E = 0.018t_0$, and (c) for $E = 0.16t_0$. (d) A transverse section of (b) along the vertical direction through the center of the hexagon, showing the profile of the LDOS and the PMF ("displacement approach"). $\tilde{R}$ marks the distance to the center.

## Quantum Transport

To calculate the quantum transport characteristics of the strained hexagon we coupled three un-strained semi-infinite metallic AC graphene nanoribbons to the sides of the ZZ hexagon where the load is applied (cf. inset of Fig. 1), thereby creating a Y-junction. There is no barrier between the metallic contacts and the central region, and the only perturbation to the electronic motion arises from the strain-induced changes in the nearest neighbor hoppings inside the hexagon. The width of the contacts coincides with the side, $L$, of the hexagon. In a multi-contact device the current in the $p$-th contact is expressed using the Landauer-Büttiker formalism as $I_p = \frac{2e^2}{h} \sum_q [T_{qp}V_p - T_{pq}V_q].$[29] With no loss of generality, a bias voltage $V_1$ is applied to contact 1, while contacts 2 and 3 are grounded. In this configuration $I_1 = \frac{2e^2}{h}[T_{21} + T_{31}]V_1$, $I_2 = -\frac{2e^2}{h}T_{21}V_1$ and $I_3 = -\frac{2e^2}{h}T_{31}V_1$, reducing the calculations to the transmission coefficient between contact 1 and 2: $T_{21}$ ($T_{31} = T_{21}$ under symmetric loading). The transmission coefficient is given by $T_{qp} = \text{Tr}[\Gamma_q G^r \Gamma_p G^a]$, where the Green's functions are $G^r = [G^a]^\dagger = [E + i\eta - H - \Sigma_1 - \Sigma_2 - \Sigma_3]^{-1}$, the coupling between the contacts and the device is $\Gamma_q = i[\Sigma_q - \Sigma_q^\dagger]$, and $\Sigma_q$ is the self energy of contact $q$, all of them calculated numerically.[30]

Fig. 4(a) shows the transmission coefficient $T_{21}$ ($= T_{31}$ under symmetric loading) as a function of the Fermi energy, $E_F$, for the Y-junction of Fig. 1. The smooth (blue) curve is the transmission in the absence of strain, and the resonant trace (black) the transmission for $\varepsilon_{\text{eff}} = 10\%$. The unstrained junction's transmission is characterized by a threshold and a broad resonance around $E/t_0 \approx 0.063$, and a set of broad resonances and anti-resonances on a smooth background as $E$ increases. The resonance at the threshold marks the fundamental mode of the hexagonal cavity which, from the geometry, is estimated to appear at $E \approx \hbar v_F (\pi/\sqrt{2}L) = 0.06 t_0$. Spatial mapping of the LDOS (not shown) at this energy confirms this.

Upon stretching, three different regions can be identified in the curve of $T_{21}(E)$ in Fig. 4(a): (i) at low energies the transmission is suppressed; (ii) at intermediate energies the transmission develops a series of regularly spaced sharp resonances; (iii) at higher energies the transmission shows unevenly spaced and rapidly oscillating peaks. To characterize these different regimes we resort to the features of the overall DOS, as well as the LDOS distribution, $N(\boldsymbol{r}, E) = (-1/\pi) \text{Im}[G^r(\boldsymbol{r}, \boldsymbol{r}; E)]$, at representative energies. The DOS is shown in the inset of Fig. 4(a) and, even though there are plenty of low energy states, only those above $E \approx 0.08 t_0$ have an appreciable signature in the transmission. To understand this absence of transmission we turn to Fig. 4(b), which plots a real-space LDOS map of a state at $E = 0.018t_0$, representative of these low energy states that have no signature in the transmission. Apart from the non-propagating LDOS accumulation at the ZZ edges, the significant LDOS amplitude is distributed within an annulus of radius $\approx 4$ nm and width $\approx 2.5$ nm. Since the LDOS does not extend to the vicinity of the contacts, revealing a small coupling between this state and the modes of the contacts, the only possibility for transmission is through tunneling. But since the spatial barrier for tunneling into this confined state is rather large ($\approx 2.5$ nm), the resonant peak in the transmission associated with this state has a vanishingly small amplitude and is not seen on the scale of Fig. 4. A transverse cut of the LDOS in (b) along the vertical direction through the hexagon center is shown in panel (d).[17] It reflects the wavefunction of a PMF-induced Landau edge state confined to the hexagonal quantum dot, analogous to the edge states in magnetic quantum dots.[28] As the energy is progressively increased, the associated states spread out, approaching the boundaries. Their coupling to the contacts increases until the tunneling-assisted conductance becomes of the order of the conductance quantum and the associated transmission resonances become visible in the black trace of Fig. 4(b).

The LDOS map in Fig. 4(c) corresponds to $E = 0.16t_0$, and typifies the behavior at higher $E_F$. It is clear that this state is completely different from the one in Fig. 4(b), as its LDOS spreads over the entire dot and is highly peaked at the center. It corresponds to a state in the $n = 1$ LL. The rapid oscillations in the transmission coefficient and DOS at that energy are also consistent with this interpretation.[31] An additional quantitative confirmation is given as follows. If the state at $E = 0.16t_0$ belongs to the $n = 1$ LL, its associated magnetic length will be $\ell_{B_s} = \sqrt{2}\hbar v_F/E_{n=1} \simeq 1.9$ nm. The energy difference between Landau edge states whose energy is between $E_{n=0}$ and $E_{n=1}$ can be estimated by dividing the LL separation by the number of edge states per LL: $\Delta E \approx (E_{n=1} - E_{n=0})\ell_{B_s}/2L_0 \approx 0.02t_0$, where the factor $2L_0/\ell_{B_s}$ corresponds to the average number of edge states between these two Landau levels.[31] Inspecting the inset of Fig. 4(a) we verify that the level spacing below $E \simeq 0.1t_0$ is indeed $\sim 0.02t_0$. Moreover, given that this quantitative estimate is consistent, we can extract the average magnetic field determining the transport behavior, which is $B_s^{\text{av}} = \hbar/e\ell_{B_s}^2 \simeq 164$ T. This value, obtained independently and solely from the transmission characteristics, expectedly corresponds to the PMF in the region of maximum LDOS for this state: from Fig. 4(d), and correcting for the overestimation in the PMFs obtained from the "displacement approach" in Fig. 3, that would be $\approx 270/1.6 = 169$ T.

Hence, transport in the strained junction is characterized by LL-assisted resonant tunneling, analogously to a magnetic quantum dot, with the novelty that here the Landau quantization arises from the strain-induced PMF, $B_s$. Due to the effective magnetic barrier, electrons injected from contact 1 can tunnel with a probability $0 < T < 1$, which is enhanced when there is significant LDOS in the contact region. The maximum tunneling probability through a localized state is $T = 1$, irrespective of the number of open channels in the contacts. This implies that between the LL $n = 0$ and $n = 1$ we expect $T_{21}^{max} = 0.5$ (0.5 because there are two symmetric exit channels). However each LL with $n \neq 0$ in graphene is doubly degenerate, and hence $T_{21}^{max} = 1$, which is consistent with the calculated transmission seen in Fig. 4(a), where $T_{21}(E = 0.126t_0) = 0.79$, for example.

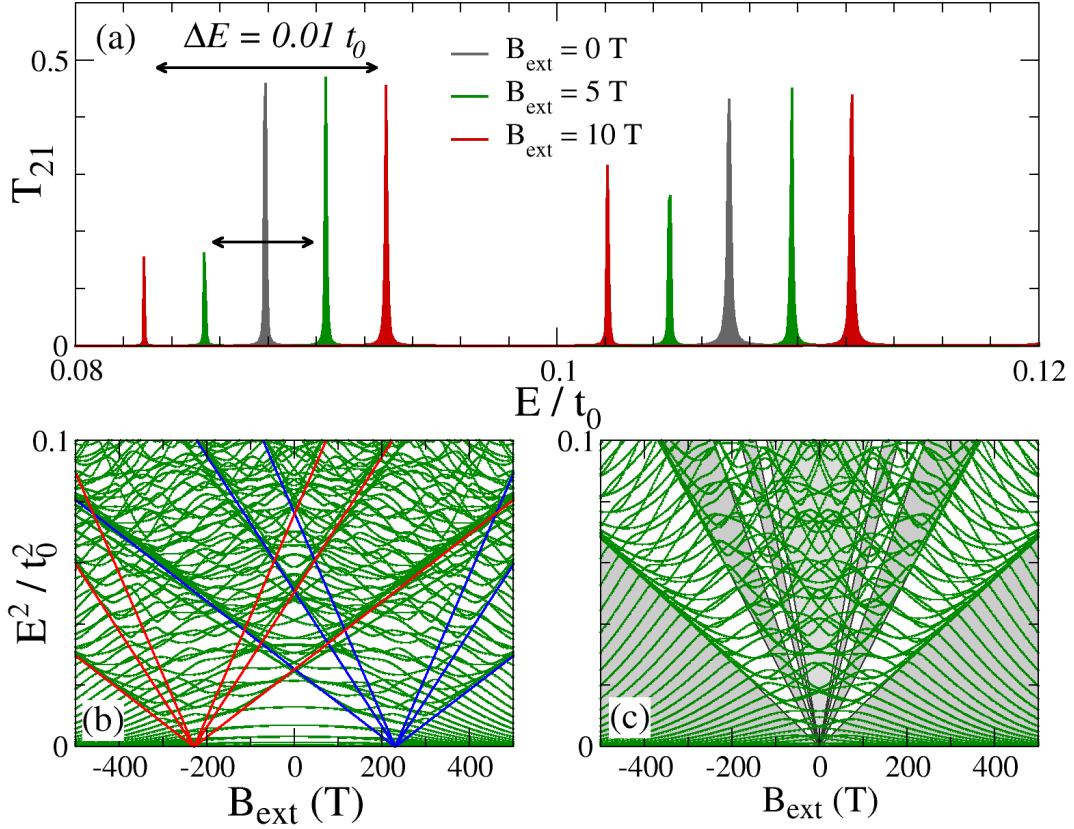

Figure 5: (Color online) (a) Detail of the splitting in the $T_{21}$ resonances under an external field $B_{ext}$, for $\varepsilon_{eff} = 10\%$. (b) Eigenenergies of the same hexagon vs $B_{ext}$, when disconnected from the contacts. (c) Likewise, but for $\varepsilon_{eff} = 0$, where LL condensation[28] is more clearly observed. In (b) and (c) straight lines mark the lowest LLs in the infinite system, and the large range of $B_{ext}$ used in the horizontal axes is to accommodate the very large PMF induced by strain.

## Valley Splitting

The strain-induced PMF does not break time-reversal symmetry (TRS) in the system, which in practice means that a low energy electron around one of the zone-edge valleys feels a PMF which is exactly the opposite to the one felt by its TRS counterpart at the other inequivalent valley. This leads to the degeneracy discussed above and, in addition, to the result that the currents associated with the two valleys exactly cancel each other. This degeneracy is lifted under a real magnetic field, $B_{\text{ext}}$, since the total field at each valley will be different: $B_{\text{ext}} \pm B_s$. The corresponding LL splitting is given by $E_n^+ - E_n^- \simeq E_n B_{\text{ext}} / B_s^{\text{av}}$. In Fig. 5(a) we show explicitly this splitting for the edge states detached from the $n = 1$ LL. Taking the values estimated above for $B_s^{\text{av}} \simeq 164\,\text{T}$, and $E_{n=1} \simeq 0.16t_0$, the expected splitting under the external field is $(E^+ - E^-)/t_0 \simeq 0.001 B_{\text{ext}}\,\text{T}^{-1}$. Direct inspection of Fig. 5(a) shows that this is indeed quantitatively verified. A different perspective over the splitting of valley degeneracy is given in Fig. 5(b), which shows the spectrum of the strained hexagon disconnected from the contacts, as a function of $B_{\text{ext}}$. When compared with the unstrained case in Fig. 5(c), one sees that the effect of the large PMF induced by strain is to split the Landau fan, which is a clear evidence of valley degeneracy breaking. Notice also that this degeneracy breaking is visibly achieved under $10\,\text{T}$, as shown in Fig. 5(a).

Another interesting consequence of breaking the valley degeneracy is that, as $B_{\text{ext}}$ increases, the edge states in one valley will shrink to a smaller radius than in Fig. 4(b), whereas the ones associated with the other will expand due to the opposite evolution of the respective magnetic lengths. Therefore, by increasing $B_{\text{ext}}$ one can spatially "expand" the edge states of one valley [cf. Fig. 4(b)] so that they start coupling more effectively with the leads. This is reflected in Fig. 5 by the asymmetry in $T_{12}$ of the split transmission resonances: the state which increases in energy under $B_{\text{ext}}$ is the one whose $\ell_B$ increases, thereby facilitating the resonant tunneling process, and displaying higher transmission than its counterpart associated with the other valley. Consequently, with an external field one can restrict the assisted tunneling to states from one or the other valley. The current path will then have a well defined chirality depending on which valley is assisting the tunneling. This suggests the possibility of exploring this chiral resonant tunneling to channel the current from lead 1 selectively to lead 2 or 3.

## Asymmetry, disorder and lattice orientation

The triaxial strain profile of Fig. 1 was chosen in this investigation as it provides a nearly optimal PMF distribution within the nanostructure.[6] However, the magnitude of the PMF will depend on the relative orientation of traction and crystal directions, implying that the magnitude of the confining effects, for example, is sensitive to that orientation. This is a general feature of strain induced PMFs in graphene. Likewise, non-symmetric triaxial tension perturbs the PMF distribution as well, which has consequences for the electronic behavior. For a perspective on this we discuss the transport behavior for different lattice orientations, as well as asymmetric tension, in the SI.[17] An analysis of the consequences of edge roughness[17] shows that the LL-assisted tunneling is sensitive to the amount of roughness at the boundaries, which is expected since the large PMF at the center of the hexagon forces the current to flow close to the boundary. In this sense, the experimental exploration of the LL-assisted tunneling described here is more straightforwardly observable in graphene structures synthesized via bottom-up microscopic approaches,[32,33] or artificial graphene structures,[26,34] where the effects of fabrication-induced disorder can be minimized. With respect to the strain symmetry, we observe that extreme deviations from symmetric traction deteriorate the LL-assisted resonant tunneling that is possible under the conditions discussed above. This arises because asymmetric strain displaces the region of strong field towards one of the boundaries. This, on the other hand, can be explored to selectively block one of the output contacts in the y-junction, allowing control over which of the two is the exit channel. A specific case is analyzed in the SI.

## Conclusion

The unparalleled elastic properties of graphene and the unusual response of its electrons to deformations captured by the PMF concept imply that nanostructures deformed with the right symmetry can behave as magnetic quantum dots. Conductance at low $E_F$ is limited to edge state assisted resonant tunneling, and the valley degeneracy can be explicitly broken under an external field, allowing control over which valley assists in the tunneling process. Since $B_s$ can easily reach hundreds of Tesla in experiments,[12,13] such small pseudomagnetic quantum dots are a viable prospect and certainly warrant further investigation.

## Acknowledgement

This work was supported by the NRF-CRP award "Novel 2D materials with tailored properties: beyond graphene" (R-144-000-295-281). ZQ and HSP acknowledge the support from NSF grant CMMI-1036460, Banco Santander, and from the Mechanical Engineering Department of Boston University. HSP further acknowledges support from a Boston University Discovery Grant.

# Resonant tunneling in graphene pseudomagnetic quantum dots

## — Supporting information —

Zenan Qi,[1] D. A. Bahamon,[2] Vitor M. Pereira,[2, *] Harold

S. Park,[1] D. K. Campbell,[3] and A. H. Castro Neto[2,3]

[1]*Department of Mechanical Engineering,*

*Boston University, Boston, MA 02215*

[2]*Graphene Research Centre & Department of Physics,*

*National University of Singapore, 2 Science Drive 3, Singapore 117542*

[3]*Department of Physics, Boston University, Boston, MA 02215, USA*

(Dated: May 8, 2013)

---

* Corresponding author: `vpereira@nus.edu.sg`

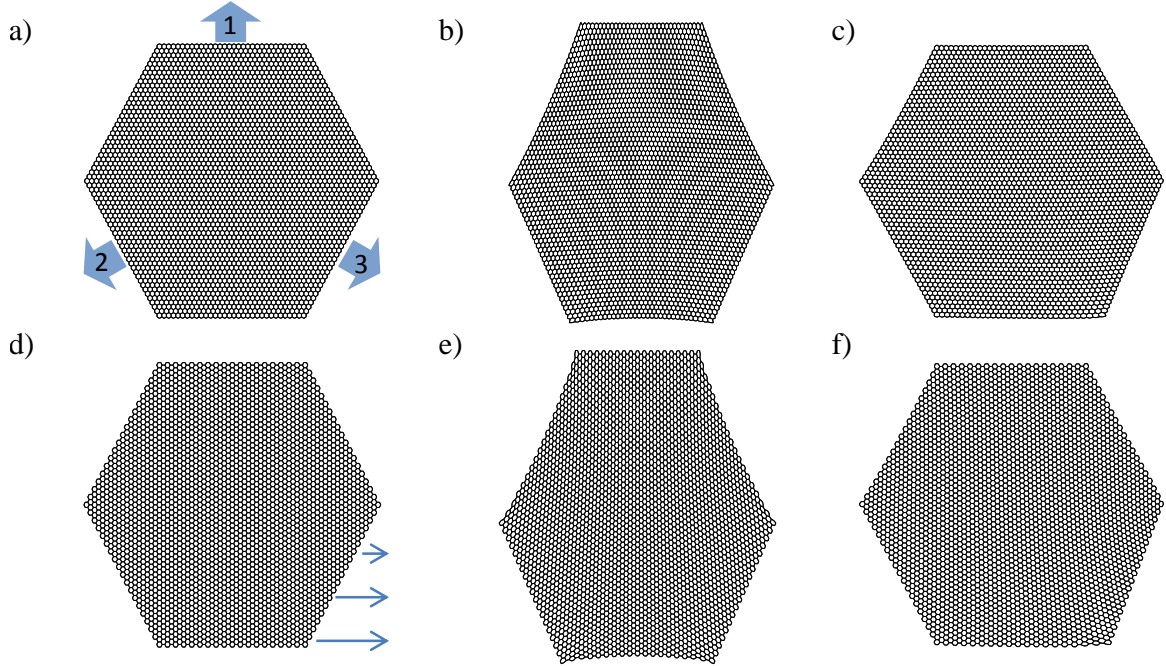

FIG. 1: (a) Hexagon (0% strain) with ZZ edges. The arrows are a schematic representation of the symmetrically applied tri-axial strain. In the transport calculations the contacts are attached to the edges under traction as well because the edge atoms are held fixed along the direction transverse to the tension, which allows us to keep the metallic contacts undeformed. (b) Deformed hexagon with ZZ edges after 10% of symmetric strain. (c) Deformed hexagon with ZZ edges after 15% of asymmetric (ramp) strain. (d) Hexagon (0% strain) with AC edges. The arrows are a schematic representation of the asymmetrical (ramp) strain setup. (e) Deformed hexagon with AC edges under 15% of symmetric strain. (f) Deformed hexagon with AC edges under 15% of asymmetric (ramp) strain.

## I. DEFORMED HEXAGONS UNDER STRAIN

The graphene hexagon was deformed via molecular mechanics using the open-source atomistic simulation package LAMMPS [1], where molecular mechanics implies deformation at 0K. The C-C interactions were modeled using the Adaptive Intermolecular Reactive Empirical Bond Order (AIREBO) potential [2]

$$E = \frac{1}{2} \sum_i \sum_{j \neq i} \left[ E_{ij}^{REBO} + E_{ij}^{LJ} + \sum_{k \neq i,j} \sum_{l \neq i,j,k} E_{kijl}^{TORSION} \right], \tag{1}$$

which consists of three parts: the 2nd generation REBO potential ($E_{ij}^{REBO}$) [3], a standard Lennard Jones potential ($E_{ij}^{LJ}$), and a many body torsion term ($E_{kijl}^{TORSION}$). As shown in Eq. (1), $E^{REBO}$ is the 2nd generation Brenner potential which dominates the short-range interactions. The Lennard-Jones $E^{LJ}$ term is only switched on for long-range interactions between 0.2 nm to the cut-off radius of the interatomic potential (0.68 nm), and is switched off for distances shorter than 0.2 nm to avoid its steep $(1/r_{ij}^{12})$ repulsive wall. Finally, the $E^{TORSION}$ term depends on dihedral angles which have little effect in this work due to the 2D nature of graphene. No periodic boundary conditions were used in this work, and the total numbers of atoms used in the zigzag and armchair sheets are 6144 and 6162, respectively. We define three edge groups (32 atoms in each group for zigzag, 38 atoms in each group for armchair) and applied displacement loading to the three edges as shown in Fig. 1(a). For asymmetric loading cases, one edge was stretched while the other two were fixed, as illustrated in Fig. 1(d).

Since molecular mechanics simulations are performed at 0K, there is no need for thermostats to control the temperature of the system, and the equilibrium (minimum energy) positions of the atoms are obtained using the conjugate gradient method with a relative energy tolerance of $10^{-7}$ eV between successive displacement increments. Specifically, the graphene hexagon is stretched by applying displacement increments of 0.01 Å perpendicular to the corresponding edges, as illustrated in Fig. 1(a). After the displacement increment is applied to the three edges, the three edges are held fixed, at which point the unconstrained atoms in the hexagon are allowed to relax to their corresponding equilibrium positions using the conjugate gradient algorithm. This stretching and relaxation loop is repeated until the desired nominal strain ($\epsilon_{eff}$) is reached. Fracture of the hexagonal sheet is observed when $\epsilon_{eff}$ goes up to ~19% for the symmetric loading depicted in Fig. 1(a). Fig. 1 shows the hexagonal quantum dots considered in this study under different loading conditions and for different lattice orientations with respect to the applied strain. The strategy used to explore deviations from the symmetric loading is shown schematically in Fig. 1(d). In this case the traction is applied only to edge 3, and in such a way that the displacement follows a ramp pattern, being maximal at one end of the edge and linearly decreasing to zero towards the opposite end. Edges 1 and 2 are held fixed Figs 1c and 1f show the actual relaxed structures after the molecular mechanics simulation under this ramp traction profile, and for 15% of nominal strain applied to the lower atoms.

## II. ARMCHAIR Y-JUNCTION

The electronic spectrum and transport characteristics of graphene nanostructures is strongly influenced by the nature of the edges. Strain-induced PMFs also depend on the relative orientation of the strain with respect to the underlying crystal directions of the graphene lattice. Here we consider a quantum dot with the same geometry and the same approximate dimensions as the one discussed in the main text, but for which the graphene lattice has been rotated so that the edges are now of the armchair (AC) type. This corresponds to a rotation of the original lattice by $\pi/2$, or any other equivalent angle.

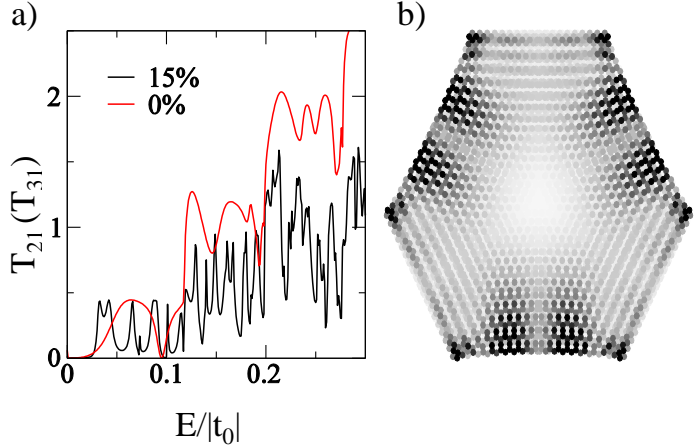

a)

b)

FIG. 2: (Color online) (a) Transmission coefficient $T_{21}(T_{31})$ *vs* Fermi Energy for a symmetrically strained AC Y-junction. (b) LDOS mapping at $E = 0.043t_0$ for the 15% symmetrically strained AC Y-junction.

The calculation of the transmission is done now by attaching three unstrained semi-infinite zig-zag (ZZ) graphene nanoribbons, which act as ideal contacts. As in the main text, the contacts are connected to the sides of the hexagon where the load is applied, creating an AC Y-junction. In Fig. 2(a) we can see the transmission coefficient for 0% strain (red). As discussed in the main text for the ZZ case, the onset of transmission in the unstrained structure is characterized by a very broad hump, that is associated with the fundamental mode of the cavity. The different nature of the AC edges manifests itself by the wider and deeper resonances and anti-resonances that develop as the energy increases, in comparison with the transmission fingerprint of the unstrained ZZ junction discussed in the main text. When strain is applied up to the nominal value of 10% the transmission coefficient mostly resembles the unstrained case, and the case of $\epsilon_{\text{eff}} = 15\%$, represented by the black curve in Fig. 2(a) is still qualitatively similar to the unstrained situation. More specifically, despite the additional structure, there are no isolated resonant peaks in contrast to the case of the ZZ junction, and transmission is never zero after the initial onset at around $E \simeq 0.02$. In clear contrast with the case analyzed in the main text, the transmission signature of this junction is not compatible with the presence of a significant pseudomagnetic field within the central region of the hexagon. Direct inspection of the real-space LDOS distribution at the transmission peaks confirms this. Fig. 2(b) represents a density plot of the LDOS for the transmission peak at $E = 0.043t_0$, revealing a LDOS distribution qualitatively similar to any resonance in the unstrained structure.

The inference that there is no significant homogeneous magnetic field within the junction from the transport fingerprint alone is compatible with the expectation for the pseudomagnetic field distribution anticipated in this case. Despite the generic relevance of the edge chirality in small graphene structures, the crucial detail in the context of generating suitable PMF distributions is the orientation of the lattice with respect to the strain directions. On the basis of the results derived in reference 4 we can expect the magnitude of the PMF near the center of the hexagon to vary with the lattice orientation as $\propto \cos(3\varphi)$, where $\varphi = 0$ corresponds to a lattice with a ZZ direction along the horizontal axis. Since the AC case studied in Fig. 2 corresponds to $\varphi = \pi/6$, $\pi/2$, etc. we expect the magnitude of the pseudomagnetic field to be mostly suppressed in the central region.

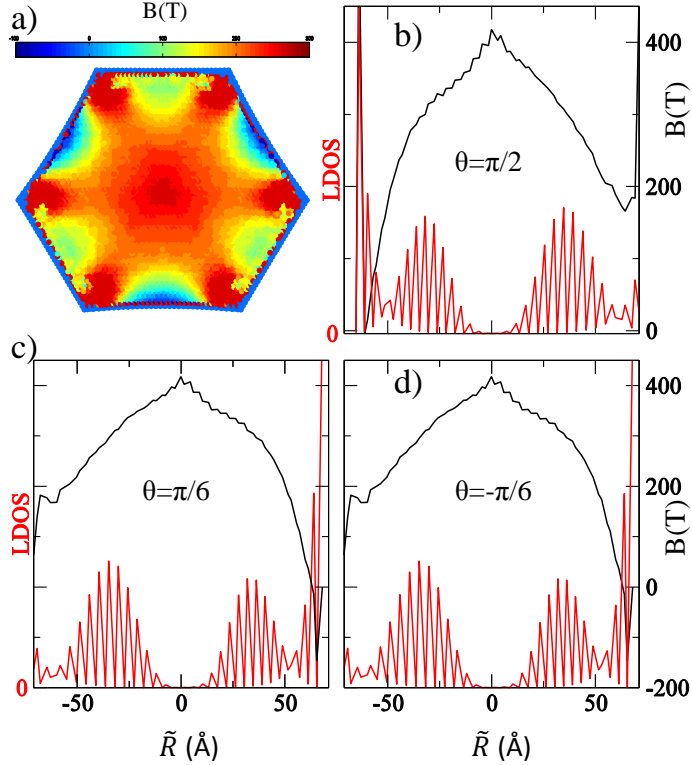

FIG. 3: (Color online) Sections of $B_s$ and LDOS for a strained ZZ Y-junction with $\epsilon_{\text{eff}} = 10\%$. Panel (a) shows a density plot of $B_s$ in the entire system obtained with the "displacement approach". In panels (b)-(d) we plot the profile of $B_s$ (black) and LDOS (red) at $E = 0.018t_0$ along the directions defined in the text: $\theta = \pi/2$, $\theta = \pi/6$, and $\theta = -\pi/6$, respectively.

## III.  PROFILE OF $B_s$ AND LDOS (ZZ JUNCTION)

In the main text we studied in detail the case of a ZZ junction under $\epsilon_{\text{eff}} = 10\%$. From the nature of the resonant transmission at low energies, and from the equidistant spacing between resonances, we extracted an average pseudomagnetic field $B_s^{\text{av}}$ determining the behavior of transmission. Moreover, we stated that the resonant transmission occurs only through the assistance of those edge states whose radius is such that they effectively couple

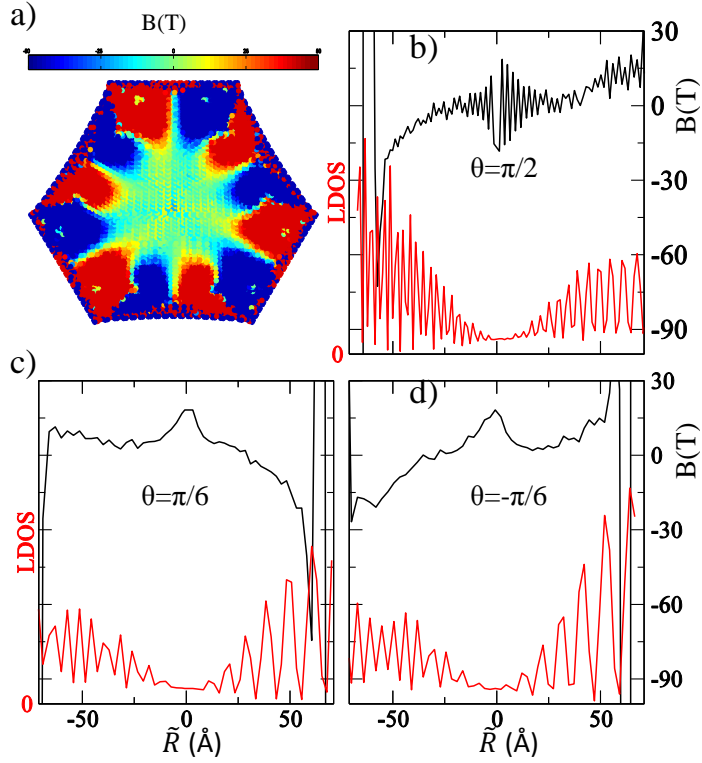

FIG. 4: Sections of $B_s$ and LDOS for a strained AC Y-junction with $\epsilon_{\text{eff}} = 15\%$. Panel (a) shows a density plot of $B_s$ in the entire system obtained with the "displacement approach". In panels (b)-(d) we plot the profile of $B_s$ (black) and LDOS (red) at $E = 0.043t_0$ along the directions defined in the text: $\theta = \pi/2$, $\theta = \pi/6$, and $\theta = -\pi/6$, respectively.

to the contacts at the border.

To clarify these points, the values of $B_s$ and LDOS at $E_1 = 0.018t_0$ in the ZZ junction with $\epsilon_{\text{eff}} = 10\%$, are extracted along transverse sections equivalent by symmetry. The origin of coordinates is set at the center of the strained hexagon, with the $x$ and $y$ axes along the conventional horizontal and vertical directions. The direction of the transverse section with respect to the horizontal axis is defined by the angle $\theta$, and the position of a point in the lattice along this section is identified by $\tilde{R} = \text{sign}(y)\sqrt{x^2 + y^2}$. We shall consider the three equivalent transverse sections along $\theta = \pi/2$ and $\pm\pi/6$. For example, contact 1 appears at $\tilde{R} \approx 74$ Å in a section taken along $\theta = \pi/2$. These three sections are chosen to confirm and highlight the isotropy of both $B_s$ and the LDOS in the interior of the structure.

The overall distribution of $B_s$ within the hexagon is shown in Fig. 3(a). The value of $B_s$ shown here is extracted using the "displacement" approach discussed in the main text. It consists in using the coordinates of the relaxed atoms directly to interpolate the strain tensor, after which the vector potential $\boldsymbol{A}_s$ is extracted. This method has the potential disadvantage of requiring a sequence of three numerical derivatives to obtain the value of $B_s$ at a given lattice point, given the relaxed atomic coordinates, and also overestimates $B_s$ at large deformations, as shown in Fig. 3 of the main text. However it is much more expedite than the mapping of the LDOS, and extraction of the local LL spectrum from the tight-binding calculation, which was the method used to plot $B_s$ in Fig. 1 of the main text. Correcting for the overestimate in magnitude discussed and shown in Fig. 3 of the main text, the distribution of $B_s$ obtained with either method in the interior of the structure is equivalent.

The values of $B_s$ and LDOS along the three sections mentioned above are plotted in Figs. 3(b)-(d), represented by the black traces. The LDOS is plotted together with the pseudomagnetic field along the three sections, represented by the red traces. First, we notice that the large peaks located at boundary $\tilde{R} = -65\,\text{Å}$ ($\theta = \pi/2$), $\tilde{R} = 68\,\text{Å}$ ($\theta = \pi/6$), and $\tilde{R} = 68\,\text{Å}$ ($\theta = -\pi/6$) are due to ZZ edge states. At the opposite boundary (where the contacts are attached) the LDOS is small, signaling that this state is well confined within the interior of the structure, and that the probability of transmission through it is small. The most interesting detail of the LDOS distribution, however, is its distribution in the interior of the structure. It is clear that the wave function does not follow local features in $B_s$, such as changes of strength or sign of $B_s$ [5]. In contrast, the LDOS intensity is almost completely confined to an annular region inside the junction, fully resembling the LDOS of an edge state in a magnetic quantum dot, as described by Lent [6]. From the LDOS profile we obtain $\ell_{B_s} \approx 2\,\text{nm}$, which corresponds to the average field $B_s^{\text{av}} = \hbar/e\ell_{B_s}^2 \simeq 164\,\text{T}$ extracted in the main text.

## IV. PROFILE OF $B_s$ AND LDOS (AC JUNCTION)

The procedure described in the previous section is applied to the analysis of the AC junction with $\epsilon_{\text{eff}} = 15\%$. As expected, the magnitude of $B_s$ is roughly zero in the interior region of the junction. Sharp features appear only around small regions near the corners and edges, where the field is strong and alternates in sign. Figs. 4(b)-(d) show the profile of $B_s$ together with the LDOS at $E = 0.043t_0$, the same energy used in Fig. 2 above.

## V. EDGE ROUGHNESS AND ASYMMETRIC STRAIN

In order to simulate the effect of edge roughness, vacancies were added with a probability of 0.4 to the edges of the two types of Y-junction. These vacancies were added in the strained electronic Hamiltonian neglecting the relaxation of local strain in the vicinity of the vacancy. This simplification should not modify our results since the main ingredient is that edge roughness reduces the transmission through pseudo-magnetic edge states (standing waves in the strain barrier) for ZZ edged Y-junctions, as can be seen in Fig. 5(a). In these structures the $B_s$ in the interior of the junction behaves like a barrier, pushing the current towards the edges. This current is effectively suppressed at low energies by the strong backscattering induced by the vacancies. This creates the gap in transmission at low energies that can be seen in Fig. 5(a). In the AC-edged hexagonal dot, since the average $B_s \approx 0$ in the center of the Y-junction, current flows easily through the central region, and there is no transmission gap.

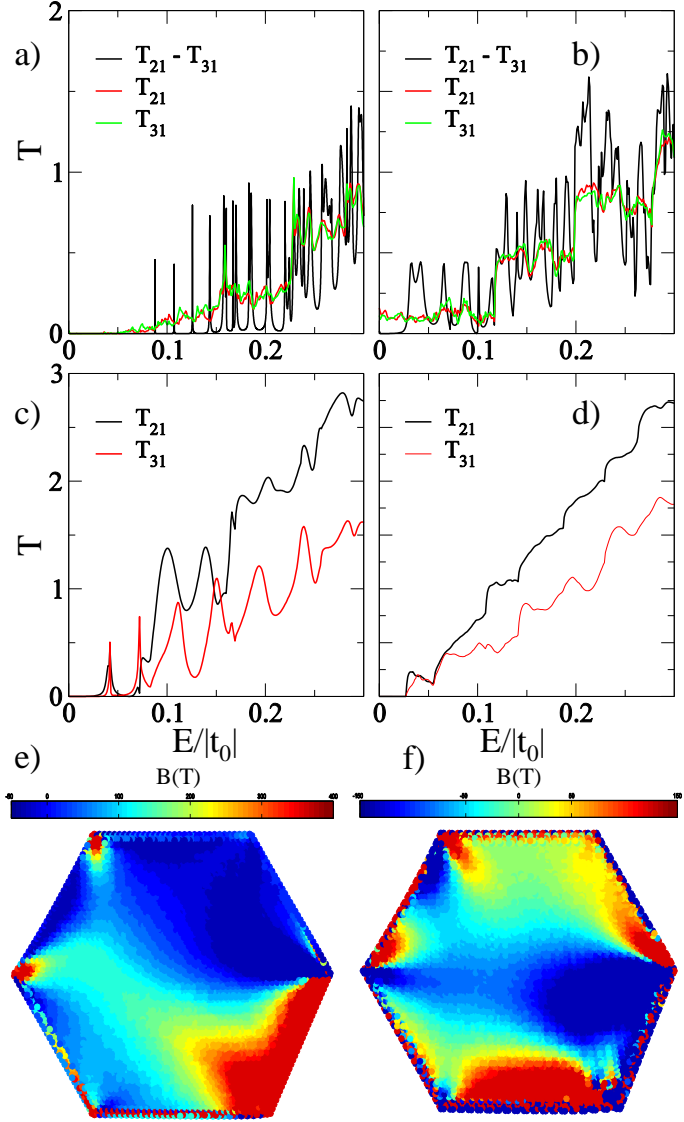

FIG. 5: Transmission coefficient for: (a) 10% strained ZZ Y-junction with and without edge disorder; (b) 15% strained AC Y-junction with and without edge disorder; (c) 15% asymmetrically strained ZZ Y-junction; (d) 15% asymmetrically strained ZZ Y-junction with edge disorder. In panels (e) and (f) we display the distribution of $B_s$ generated by the asymmetric traction illustrated in Fig. 1(d) for the case of, respectively, the ZZ and AC edged hexagon nanostructures considered in this work.

To address the problem of asymmetry in the traction (and consequently in the overall strain distribution), we consider the extremely asymmetric situations illustrated in Figs. 1(c) and (f). The transmission data is shown in Fig. 5(c), for a structure where a ramp displacement is applied only to the hexagon side where the third contact is attached. An asymmetric strain pattern ensures that $T_{21} \neq T_{31}$, and can potentially be explored to channel the current between specific pairs of contacts by suitable asymmetric traction conditions. The ramp strain considered here creates a $B_s$ that is not uniform in the center of the hexagon, but has a strong maximum in the vicinity of the third contact. Although the values of $B_s$ in that region are large ($\ell_{B_s} < L_0$), and of the same order of magnitude as the ones found in the symmetric junction, the transmission and LDOS signatures are rather different from the signatures of a symmetrically strained hexagon. In particular the resonant peaks in $T_{31}$ at $E = 0.042t_0$ and $E = 0.072t_0$ are due to states having a LDOS distribution of two distorted standing waves, rather than the magnetic edge state profiles seen in the symmetric case in Fig. 3. The reason for the different behavior is mostly due to the non-uniform nature of $B_s$ in the interior of the system, in comparison with the symmetrically strained situation. This means that the electrons don't feel a quantum dot with a nearly constant magnetic field everywhere in this case, but instead are scattered from the regions of higher field.

In essence, this extreme asymmetric case results in a distribution of $B_s$ that acts as a barrier for current flow only in certain regions inside the Y-junction. Since that barrier is higher in the region of contact 3 the current is scattered to contact 2 and, consequently, $T_{31} < T_{21}$. This imbalance in $T_{31}$ vs $T_{21}$ becomes even more evident at higher energies. Inside the ZZ Y-junction, for low energies, the current flows through the regions of low $B_s$, as shown in more detail the next section of this supporting information.

The effect of edge disorder in the asymmetric ZZ junction is presented in Fig. 5(d), where we see that the effect is not as marked as in the symmetric case shown in Fig. 5(a). This is consistent with the above description of the transmission process in this case, whereby electronic current flows through the large portions of the hexagon that are not under a significant $B_s$.

## VI. $B_s$ AS A BARRIER

The current between neighboring sites $m$ and $n$ can be expressed as [7, 8]

$$I_{mn} = \frac{2e}{h} \int_{-\infty}^{+\infty} dE [t_{nm} G_{mn}^< - t_{mn} G_{nm}^<], \qquad (2)$$

where the lesser Green's function in the absence of interactions can be written as $G^<(E) = G^r(E)[\Gamma_1 f_1 + \Gamma_2 f_2 + \Gamma_3 f_3]G^a(E)$. $f_{1(2)(3)}$ is the Fermi distribution in the respective electrodes. We mapped the current density for different energies for the ZZ and AC Y-junction with symmetric and asymmetric strain. For reference, we shown in Fig. 6 the current maps for some of the cases discussed in the main text, as well as in the previous sections of this supporting information.

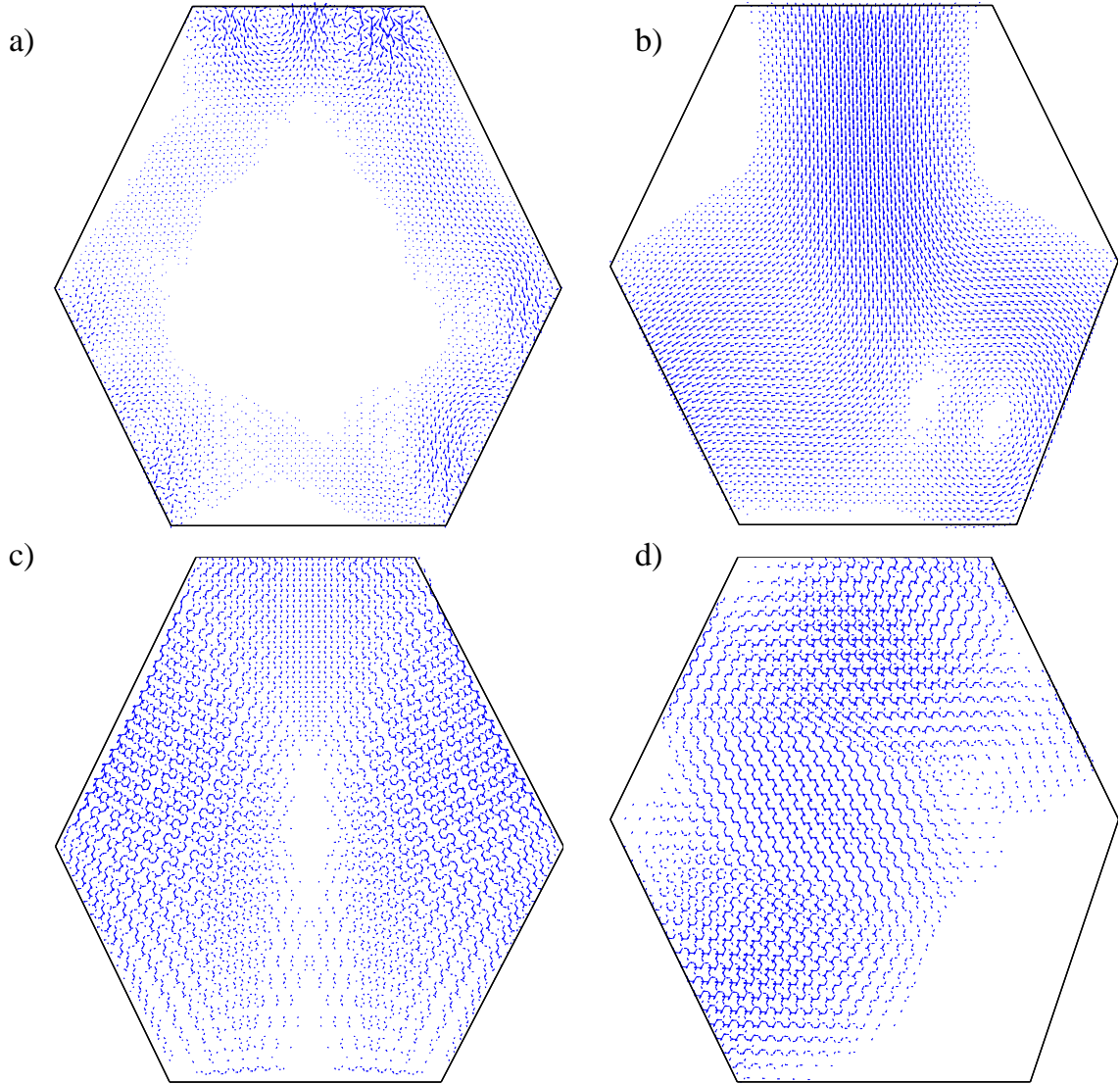

FIG. 6: Current density mapping in real space for selected transmission features discussed in the main text and above: (a) $E = 0.126t_0$ for the symmetric ZZ Y-junction with $\epsilon_{\text{eff}} = 10\%$; (b) for the asymmetrically strained ZZ Y-junction with $\epsilon_{\text{eff}} = 15\%$ at $E = 0.1t_0$; (c) at $E = 0.033t_0$ for the symmetric AC Y-junction with $\epsilon_{\text{eff}} = 15\%$; (d) for the asymmetrically strained AC Y-junction at $E = 0.095t_0$. In all plots, the length of the arrow is proportional to the value of the density current in that point.

The case plotted in Fig. 6(a) pertains to a symmetrically strained ZZ Y-junction, at the energy $E = 0.126t_0$ that corresponds to one of the isolated transmission resonances. Since $B_s$ is strong in most of the interior region the current path exhibits the intuitively expected behavior by flowing through the regions of smallest field towards the edges. Due to the microscopic details of $B_s$ the current density distribution is not perfectly symmetric between contacts 1–2 and contacts 1–3. A higher density of current flows between 1–3, and part of it is scattered from contact 3 to contact 2 ensuring final transmissions of $T_{21}(0.126t_0) = 0.31 \approx T_{31}(0.126t_0) = 0.32$.

For the asymmetrically strained ZZ Y-junction in Fig. 6(b) the bulk of the current flows directly through the center of the junction, exiting predominantly via contact 2. Most of the current near contact 3 is scattered towards 2, since in the asymmetric ZZ case the magnetic barrier induced by $B_s$ is displaced to the vicinity of 3. This explains the quantitative imbalance in the respective transmissions: $T_{21}(0.1t_0) = 1.37 > T_{31}(0.1t_0) = 0.47$ [cfr. Fig. 5(c). The qualitative picture is similar for the AC Y-junction asymmetrically strained by 15% in Fig. 6(d). The main point is that under asymmetric traction conditions the distribution of $B_s$ is no longer nearly homogeneous in the central region, and a strong maximum appears towards one of the pulling arms. This restricts the magnetic barrier to a particular portion of the system, but does not lead to the Landau level assisted tunneling resonances seen in the symmetric ZZ case, and discussed in the main text. The asymmetric cases can be understood intuitively by considering the regions of strong $B_s$ as barriers that divert the electronic current, and lead to an asymmetry in the conductance measured between contacts 1–2 and 1–3.

For completeness in Fig. 6(c) we show the current density in a 15% symmetrically strained AC Y-junction, extracted at a maximum in the transmission at $E = 0.033t_0$.

---